\documentstyle[multicol,aps,prl,psfig]{revtex} 

\renewcommand{\narrowtext}{\begin{multicols}{2} \global\columnwidth20.5pc}
\renewcommand{\widetext}{\end{multicols} \global\columnwidth42.5pc}
\def\Z{{\mathcal{Z}}}
\def\csl#1{\slash\!\!\!\! #1}
\def\ssl#1{\slash\!\!\! #1}
\def\tr{{\mathrm{tr}}}
\def\fc#1#2{\frac{#1}{#2}}
\def\c{{\sc c}}
\def\d{{\sc dim}}
\def\nn{\nonumber}
\begin{document}

\title{Level-Rank Duality in Kazama-Suzuki Models}
\author{Tibra \ Ali$^\dagger$} 
\address{D.A.M.T.P., C.M.S., Wilberforce Road, Cambridge University, 
Cambridge CB3 0WA, England} 
\date{preprint DAMTP-2002-11; hep-th/0201214}

\maketitle

\begin{abstract}
We give a path-integral proof of level-rank duality in
Kazama-Suzuki models for world-sheets of spherical topology.
\end{abstract}

\narrowtext

Since the discovery of nonabelian bosonization in two dimensions
\cite{W,PW} we have learned quite a bit about the conformal
structure of string theory and two dimensional statistical field
theory. An important class of supersymmetric conformal field theories
was discovered by Kazama and Suzuki [KS] \cite{KS}. These models are
believed to be that subset of the $N=1$ supersymmetric Goddard-Kent-Olive [GKO]
\cite{GKO} coset models which admit $N=2$ superconformal symmetry. Therefore, they give
explicit integrable conformal field theory realization of target
spaces with space-time supersymmetry. Since in a lagrangian formulation
coset models correspond to the gauging a WZW model by
$g(z,\bar{z})\rightarrow h(z,\bar{z})g(z,\bar{z})h^{-1}(z,\bar{z})$
they describe, in string theory, space(times) with black holes \cite{W2,L,BCR}. Some of the KS models have been
conjectured to enjoy the following duality property \cite{KS}:
\begin{eqnarray}
\left[SU(M+N)/(SU(M)\times SU(N) \times U(1))\right]_{k}\nonumber \\
\approx\left[SU(k+M)/(SU(M)\times SU(k) \times U(1))\right]_{N} \label{eq:SU}
\end{eqnarray}
and similarly
\begin{eqnarray}
\left[SO(N+2)/(SO(N)\times SO(2))\right]_{k}\nonumber \\
\approx\left[SO(k+2)/(SO(k)\times SO(2))\right]_{N}. \label{eq:SO}
\end{eqnarray} 
There is now a considerable body of work on this duality, especially
in the operator approach \cite{NS1,FS} but a path-integral formulation of the duality has been lacking. It is
desirable for aesthetic as well as string theoretic reasons to have a
path-integral analysis of this intriguing duality. The purpose of this
letter is to supply such a proof for world-sheet of spherical
topology. We shall give our results for the case
$\left[SU(N+1)/(SU(N)\times U(1))\right]_{k}\approx
\left[SU(k+1)/(SU(k)\times U(1))\right]_{N}$ in detail but the method can be
generalized for all the models involved.
\par
It has been known for some time that an $SU(N)$ WZW model at level $k$
describes $N$ Dirac {\it parafermions} of order $k$
\cite{AB,RS}:
\begin{eqnarray}
\Z_1=\int[dg]\;e^{-kI_{SU(N)}(g)}=\int[d\psi][d\bar{\psi}][dA]\;e^{-I_{F}(\psi,A)}
\label{eq:AB}
\end{eqnarray}
where $I_{SU(N)}$ is the level-$1$ WZW action for the
group $SU(N)$ and the fermionic action is given by $I_{F}(\psi,A)=\fc{1}{2\pi}\int d^2z
\;i\bar{\psi}^{i}_{a}{\ssl{\partial}}\psi^{i}_{a}+
\bar{\psi}^{i}_{a}{\csl{A}}^{I}\lambda^{I}_{ij}\psi^{j}_{a}$ with
$i,j=1,\cdots, k$, $I=1,\cdots, k^2$ and  $a=1,\cdots, N$.  Because of
its equivalence to the constrained fermionic system the $SU(N)$ WZW
model at level $k$ can also be expressed in terms of a negative level
WZW model, free fermions, a free boson and ghosts \cite{NS2}:
\begin{eqnarray}
\Z_1&=&\int [d\bar{\psi}][d\psi][d\hat{g}][d\Phi]
\det{\partial_+}\det{\partial_-}\;e^{-I_B(\Phi)} e^{-I_{F}(\psi,0)}\nonumber \\
&\times&
e^{-[-N-2k]I_{SU(k)}(\hat{g})}.\label{eq:NS}
\end{eqnarray}
Where $I_B(\Phi)$ is the action of a free boson. Negative level WZW
models are non-unitary but since they arise here
with ghosts, it is believed that a BRST cohomology structure ensures
their unitarity. This
point has been explored in \cite{KS2}.
\par
There are various approaches to obtaining a supersymmetric extension
of a gauged-WZW model on $G/H$ \cite{W3,S,T}. The one that we follow here is the one by
Witten \cite{W3}. In this approach the supersymmetric gauged-WZW
action is given by
\begin{eqnarray}
I(g,A,\psi)&=&I(g,A) \nonumber \\
&+&\fc{i}{4\pi}\int d^2z\;
\tr(\psi_{+}D_-\psi_{+}+\psi_{-}D_+\psi_{-}).
\label{eq:Witten}
\end{eqnarray}
Here $I(g,A)$ is the bosonic action for a gauged-WZW model and the $\psi_\pm$ are Weyl fermions which belong to the complexified
orthogonal complement of the Lie algebra ${\mathcal H}$ and they are minimally
coupled to the gauge fields $A_\pm$. The condition that the $N=1$
superconformal algebra can be extended to $N=2$ superconformal algebra
can be translated to the condition that the group coset $G/H$ is
K\"ahler. Eq. (\ref{eq:Witten}) can then be written in a suitable form
to reflect that structure \cite{W3}.
\par
We are interested in the partition function for the KS models as
gauged-WZW models:
\begin{eqnarray}
\Z_2&=&\int [dg][dA_+][dA_-][d\psi_+][d\psi_-]\;e^{-kI(g,A,\psi)} \nonumber \\
&=&\int
[dg][dh][d\tilde{h}][d\psi_+][d\psi_-]\det{D_+}\det{D_-}\nonumber \\
&\times&
e^{-kI_{G}(g)}e^{kI_{H}(h^{-1}\tilde{h})}e^{\fc{1}{2}(c_{G}-c_{H})I_{H}(h^{-1}\tilde{h})}e^{I_{F}(\psi,0)}. \label{eq:GWZW}
\end{eqnarray}
Where we have made the change of variables $A_-=\partial_-
\tilde{h}\tilde{h}^{-1},\;A_+=\partial_+ h \;h^{-1}$ \cite{KS2}. $c_G$ ($c_H$) is
the quadratic Casimir operator of the group $G$ ($H$) in the adjoint representation. In the above
expression the level
$-\fc{1}{2}(c_G - c_H)$ WZW model arises from ``rotating away'' the gauge
coupling to the fermions \cite{T}. We can now use
the following identity \cite{PW}
\begin{equation}
\det{D_+}\det{D_-}=e^{c_H I_{H}(h^{-1}\tilde{h})}\det{\partial_+}\det{\partial_-}
\end{equation}
and rewrite eq. (\ref{eq:GWZW}) in a gauge fixed form (which does not
introduce any new ghosts):
\begin{eqnarray}
\Z_2&=&\int
[dg][dh][d\psi_+][d\psi_-]\det{\partial_+}\det{\partial_-}\nonumber \\
&\times& e^{-kI_{G}(g)}e^{(k+\fc{1}{2}(c_G-c_H)+c_H)I_{H}(h)}
e^{-I_F(\psi,0)}.
\end{eqnarray}
We now specialize to the case where $G=SU(N+1)$ and $H=SU(N)\times
U(1)$. Using eqs. (\ref{eq:AB}) \& (\ref{eq:NS}) this can be written as
\begin{eqnarray}
\Z_2&=&\int [d\hat{g}][d\psi'_-][d\psi'_+][d\Phi']\det{\partial'_+}\det{\partial'_-}\nonumber
\\
&\times& 
e^{-I_B(\Phi')} e^{-[-N-1-2k]I_{SU(k)}(\hat{g})} e^{-I_F(\psi',0)} \nonumber \\
&\times& [dh][d\Phi][d\psi_+][d\psi_-]
\det{\partial_+}\det{\partial_-} \nonumber \\
&\times& e^{-I_B(\Phi)} e^{(k+1+2N)I_{SU(N)}(h)} e^{-I_F(\psi,0)}. \label{eq:symmetric}
\end{eqnarray}
In the above expression primed spinors correspond to free Dirac
spinors with global $SU(N+1)\times SU(k)\times U(1)$ symmetry. The $Nk+k$ free
Dirac spinors $\psi'$
can be combined with the original $2N$ Weyl spinors $\psi$ to give $Nk+N$ free
Dirac spinors $\chi'$ and $2k$ free Weyl spinors $\chi$. Having
rearranged the spinors in this way eqs. (\ref{eq:AB}) \& (\ref{eq:NS}) can be applied to
the level $-(k+1+2N)$ WZW model on $SU(N)\times U(1)$ to yield
\begin{eqnarray}
\Z_2&=&\int[d\hat{h}][d\hat{g}][d\chi_+][d\chi_-][d\Phi]\det{\partial'_+}\det{\partial'_-}\nonumber
\\
&\times& e^{-I_B(\Phi)}e^{-N I_{SU(k+1)}(\hat{h})} \nonumber \\
&\times& e^{-[-N-1-2k]I_{SU(k)}(\hat{g})} e^{-I_F(\chi,0)}.
\end{eqnarray}
This is the gauge-fixed partition function for a $\left[SU(k+1)/(SU(k)\times
U(1))\right]_N$ gauged-WZW model with $\hat{h}\in SU(k+1)$ and
$\hat{g}\in SU(k)$. The central charge can be easily obtained from
eq.(\ref{eq:symmetric}) which is symmetric in $k$ \& $N$. In an obvious notation the
central charge is given by
\begin{eqnarray}
\c_{\sc tot} &=& \c(SU(k),-N-1-2k)\nn \\
&+&\c(SU(N),-k-1-2N)\nonumber \\
&+& \c'_{ghost}+\c_{ghost}+2\c_{\Phi}+\c_{\psi'}+\c_{\psi}
\end{eqnarray}
where $\c(G,k)=\fc{2k\;\d(G)}{2k+c_{G}}$ is the standard central
charge for a level-$k$ WZW model on $G$,
$\c'_{ghost}=-2k^2$, $\c_{ghost}=-2N^2$, ${\c_{\Phi}=1}$,
${\c_{\psi'}=Nk+k}$ and ${\c_{\psi}=N}$, which give
\begin{equation}
\c_{\sc tot}=\fc{3kN}{k+N+1} \label{eq:charge}
\end{equation}
which is the anomaly found by KS \cite{KS}.
\par
The above proof can be extended easily to all the models involved
in (\ref{eq:SU}). Following the same set of steps as before we
arrive at an expression for the
partition function
\begin{eqnarray}
\Z_3&=&\int
[d\hat{g}][dh_1][dh_2][d\psi_+][d\psi_-][d\psi'_+][d\psi'_-][d\Phi][d\Phi'] \nn \\
&\times& e^{(M+N+2k)I_{SU(k)}(\hat{g})} e^{-I_B(\Phi')}\det{\partial'_{+}}\det{\partial'_{-}} \nn \\
&\times& e^{(k+N+2M)I_{SU(M)}(h_1)}\det{\partial_{1+}}\det{\partial_{1-}}\nn \\
&\times& e^{(k+M+2N)I_{SU(N)}(h_2)} e^{-I_B(\Phi)}
\det{\partial_{2+}}\det{\partial_{2-}} \nn \\
&\times& e^{-I_F(\psi',0)}e^{-I_F(\psi,0)} \label{eq:symmetric2}
\end{eqnarray}
where we have written the ghosts on the same line as the corresponding
WZW model. This expression is completely symmetric in $k$, $M$ and $N$
and it can be easily verified that conformal anomaly agrees with
\cite{KS}. By using analogous identities \cite{RS2} for orthogonal
groups this analysis may be extended to (\ref{eq:SO}).
\par
Eqs. (\ref{eq:symmetric}) \& (\ref{eq:symmetric2}) lay bare the structure of the level-rank
duality in a concise form. It can be used to study the precise map between the
correlation functions of the dual theories, modular invariance and
coupling to gravity. 
Also interesting is the fact that this duality
connects a string theory vacuum in the semiclassical limit (for $k$,
$N$ or
$M\rightarrow \infty$) to a string theory without a geometric
interpretation (and with infinite nav\"{\i}e ``dimensions''.) 
These issues
are under investigation \cite{A}.
\begin{center}
{\small \sc Acknowledgements}
\end{center}
It is a pleasure to thank Malcolm Perry \& Adam Ritz for helpful
discussions. I am also grateful to J\"urgen Fuchs \& Christoph
Schweigert for pointing out additional references and insightful
comments for futher investigations.
\\
\\
{\small$^{\dagger}$E-Mail: {\tt T.Ali@damtp.cam.ac.uk}}

\end{multicols}
\end{document}